# Critical Infrastructure Protection: having SIEM technology cope with network heterogeneity

Gianfranco Cerullo, Valerio Formicola, Pietro Iamiglio, Luigi Sgaglione
Department of Engineering
University of Naples "Parthenope"
Naples, Italy
{gianfranco.cerullo, valerio.formicola, pietro.iamiglio, luigi.sgaglione}@uniparthenope.it

*Abstract*— **Coordinated and targeted cyber-attacks to Critical Infrastructures (CIs) are becoming more and more frequent and sophisticated. This is due to: i) the recent technology shift towards Commercial Off-The-Shelf (COTS) products, and ii) new economical and socio-political motivations. In this paper, we discuss some of the most relevant security issues resulting from the adoption in CIs of heterogeneous network infrastructures (specifically combining wireless and IP trunks), and suggest techniques to detect, as well as to counter/mitigate attacks. We claim that techniques such as those we propose here should be integrated in future SIEM (Security Information and Event Management) solutions, and we discuss how we have done so in the EC-funded MASSIF project, with respect to a real-world CI scenario, specifically a distributed system for power grid monitoring.**

## I. Introduction

Many daily operations currently rely on services provided by systems that are referred to as Critical Infrastructures (CIs). Typical examples of CIs are: the electric grid (including the emerging technologies known as smart grids), oil and natural gas production and distribution infrastructures, transportation systems, and water supply networks. In the past, CIs consisted of components which were physically and logically separated, with well-defined interfaces and relatively little interdependence. As market needs evolved towards more efficient and innovative services, with more demanding user requirements and customer expectations, utility providers had to develop CIs where Information Technology (IT) gained more and more importance. Current CIs include a number of "cyber components", which collectively account for a significant fraction of the overall system. These cyber components are currently connected through heterogeneous networks, and they are in charge of activities that are vital to the CI, and ultimately to the society. In particular, independently of the specific characteristics of the service they provide and of the deployment context, virtually all CIs rely - to a significant extent - for their operation on the existence and on the dependability of the underlying network infrastructure. Unfortunately, the increasing interconnectivity, complexity and heterogeneity of the communication networks and systems used to connect such cyber components also increase their level of vulnerability. Furthermore, the progressive disuse of dedicated communication architectures and proprietary networked components, together with the growing adoption of IP-based solutions, exposes CIs to cyber-attacks coming from the Internet [1]. In conclusion, current CIs are characterized by a vulnerability level similar to the one of other systems which are connected to the Internet, and it will be even more so in the future. This fact is particularly scaring if one considers the dramatic socio-economic impact that CI failures can have. For the above described reasons, developing mechanisms for protecting the underlying network infrastructure of a CI from attacks and failures, in order to ensure secure end-to-end transmission of information is of paramount importance. Currently, the main tool for the protection of complex distributed systems is Security Information and Event Management (SIEM) technology. Regrettably, current SIEM technology has limited ability to cope with network heterogeneity.

This paper makes three important contributions. First, we provide a detailed treatment of the security issues resulting from the adoption in CIs of heterogeneous and novel network solutions, and specifically: Wireless Sensor Networks and QoS-enabled IP connections. Second, we propose techniques for enhancing current SIEM technology, by improving its capability of detecting and mitigating attacks targeting the heterogeneous network infrastructure of a CI. Third, we show how these techniques can be implemented with respect to a challenging real world scenario, specifically the Wide Area Monitoring System of a power grid.

The rest of the paper is organized as follows. In section II, we provide an overview of the current SIEM offer, and highlight its main limitations. In section III, we present a typical architecture of a SIEM system. In Section IV we discuss how SIEM technology might detect and mitigate attacks targeting heterogeneous network infrastructures, with respect to the case study of a power grid. Finally Section V closes the paper with some final remarks.

## II. SIEM TECHNOLOGY: STATE OF THE ART AND GAP ANALYSIS

SIEM products emerged ten years ago as a solution to the problem of data overload. They are essentially a combination of previously unbundled security management services. A SIEM solution effectively combines elements of Security Information Management (SIM) with Security Event Management (SEM). One of the main features of these solutions is their advanced log management capabilities. Log management is the process of dealing with large volumes of computer generated log messages. The key issues with log management tend to be the sheer volume of the log data and the diversity of the logs. A SIEM product typically correlates, analyzes, and reports information from a variety of data sources such as network devices, identity management devices, access management devices and operating systems. The end result is a holistic view of IT security. There are a number of leading providers in this area, most notably HP-ArcSight, EMC-RSA, and IBM (Q1 Labs) [2]. HP-ArcSight is viewed by most as the market leader in this area with their Enterprise Security Manager (ESM), which functions as an integrated product suite for collecting, analyzing, and assessing security and risk information. Q1 Labs have experienced a period of rapid growth through their QRadar appliances due to their targeting of large enterprises. IBM's QRadar SIEM appliances provide log management, event management, reporting and behavioral analysis for networks and applications. QRadar can be deployed as an all-in-one solution for smaller environments, or it can be horizontally scaled in larger environments using specialized event collection, processing and console appliances. A distinguishing characteristic of the technology is the collection and processing of NetFlow data, deep packet inspection (DPI) and behavior analysis for all supported event sources. RSA, The Security Division of EMC (with enVision, NetWitness and Security Analytics) has one of the largest SIEM installed bases. However, during 2012, competitors continued to identify RSA enVision as the most frequently displaced SIEM technology. Customers report ad hoc query and report performance issues with the enVision platform as primary reasons for considering replacement. RSA has almost completed the transition from enVision to RSA Security Analytics, which incorporates traditional SIEM functionality and is based on the NetWitness platform. RSA Security Analytics provides log and full packet data capture, threat detection, basic security monitoring and basic security analytics. The most widely used Open Source SIEM is Open Source Security Information Management (OSSIM), by AlienVault, released under the GPL license. The main objective of OSSIM is correlating alerts issued by already available security tools to increase precision and recall of security breach detection. OSSIM provides integration, management, and visualization of events of more than thirty open source security tools, and allows the integration of new security devices and applications.

We performed a thorough analysis of the available SIEM products and claim that current SIEM technology has two main limitations:

SIEM scope is mostly limited to infrastructure. This results in the inability to interpret events and incidents from other layers - such as the service view, or the business impact view -and/or from the viewpoint of the service itself. The applicability and expressiveness of SIEM should be extended from the infrastructure domain, where it is mostly confined today, to a multi-domain view involving high-level processes and services, in order to perform security-related event processing and monitoring at the service level.

Even at the infrastructure level, current SIEM technology has a limited capability of dealing with network heterogeneity. It is worth emphasizing that network heterogeneity will affect CIs more and more in the future, especially in the case of complex CIs, where Commercial Off-The-Shelf (COTS) components are largely used and communication relies on heterogeneous network technologies. Cross-layer correlation of network-related security events is key to provide effective protection of current and future CIs.

By proposing techniques for making SIEM technology capable of coping with network heterogeneity at the infrastructure level, we directly address issue 2 (since we improve the capability of SIEM technology of detecting and mitigating attacks targeting the heterogeneous network infrastructure of a CI), and indirectly address issue 1 (since we pave the way to multi-level/multi-domain security event processing, a key pre-requisite for improving the performance and the effectiveness of SIEM attack detection features).

## III. SIEM ARCHITECTURE

A typical SIEM is composed of six separate parts or processes. These parts are the source device, log collection, parsing/normalization of the logs, the rule engine, log storage, and event monitoring and retrieval [3]. The first component of a SIEM architecture is the source device that produces information and feeds such information into the SIEM. The source device can be an actual network device, such as a router, switch, or some type of server, but it can also be logs from an application or just about any other data that can be acquired, and then stored and processed in the SIEM. Reports on normal or suspicious activities are generated by applications (Web Server, DHCP, DNS, etc.), appliances (router, switch, etc.) or operating systems (Unix, Mac OS, Windows, etc.). Typically, most of the reports are logs in application specific format. Log Collection component is responsible for gathering logs from source devices. Two fundamental collection methods are used to retrieve logs from the source devices: the push method, i.e. the source device sends its logs to the SIEM system, and the pull method, i.e. the SIEM system reaches out and retrieves the logs from the source device. Parsing and Normalization component is in charge of parsing the information contained in the logs and translating it from the native format to a format manageable by the SIEM engine. Moreover, the Normalization component is in charge of filling the reports with extra information required during the correlation process. Rule and Correlation Engines trigger alerts and produce detailed reports; they work on the huge amount of logs generated by the source devices. The Rule Engine raises the alert in case specific conditions are detected in the logs, while the Correlation Engine correlates information in order to produce a more concise and precise report. The correlation task consists of matching multiple standard events from different sources into a single correlated event. This task is performed in order to make incident response procedure more immediate and effective because it generates a single event to be processed by the downstream SIEM components instead of the multiple events coming from various source devices. Log Storage component stores logs for retention purposes and historical queries; usually the storage is based on a database, a plain text file or binary data. Monitoring component allows for the interaction between the SIEM user and the SIEM framework. Interactions include report visualization, incident handling, policy and rule creation, database querying, asset analysis, vulnerability view, event drilling down, and system maintenance.

## IV. IMPROVING SIEM ABILITY TO DEAL WITH NETWORK HETEROGENEITY: THE POWER GRID CASE STUDY

In this section, we show how an enhanced SIEM system, such as the one developed in the context of the MASSIF project [4], can be used to detect attacks exploiting specific vulnerabilities affecting the network trunks which compose the communication layer supporting a Critical Infrastructure. The basic idea is to collect information at several architectural levels and from different domains, using multiple security probes, which are deployed as a distributed architecture, to perform sophisticated correlation analysis of attack symptoms. In order to effectively assess the security status of the CI being protected, the results of the monitoring activities performed at different observation points need to be correlated. Such observation points are distributed throughout the network infrastructure as well as throughout the system to be protected. The more diverse the information sources and the processing methods, the more effective the correlation process. The deployment of probes at different observation points located in the communication layer and at different architectural levels (network level, operating system level, application level, etc.), allows to fulfill the requirement of diversity of the information sources. By exploiting information diversity, it is possible to improve the accuracy of the detection process, as well as to implement diagnostic capabilities.

In order to process and correlate the huge amount of heterogeneous monitoring data the proposed SIEM solution relies on the Complex Event Processing (CEP) technology. We present our solution with respect to a case study that well illustrates how the deployment of heterogeneous networking technologies impacts the security of a CI, specifically a Wide Area Monitoring System for power grid. In the following, we provide a brief description of a typical power grid infrastructure. An Electrical Power System (EPS) (or Electrical System) includes all the devices necessary to produce and to transport electric power from production plants to final users. The EPS is normally organized in islands, named grids or power grids, interconnected for increased reliability and availability. Within an island everything is synchronous, that is operates at the same frequency, while different islands can operate at different frequencies. Phasor Measurement Units (PMUs) are devices that use Global Positioning System (GPS) signals as a common time source and measure power system quantities at different locations across a wide-area system at the same instant in time. A reference architecture for power grid monitoring is composed of PMUs disseminated throughout the grid, Phasor Data Concentrators (PDCs) collecting and correlating measurement data from the PMUs, and a SuperPDC, which is a central PDC that gathers measurement data from all remote PDCs and PMUs and makes it available to a visualization station. An accurate description of the power grid monitoring architecture is provided in [5], whereas more details about security issues in power grid monitoring infrastructure can be found in [6]. The communication infrastructure supporting a power grid includes several network technologies, such as Wireless Sensor Networks (WSNs), IP-based wired network, and satellite network. Due to the low cost and high function, wireless sensors have been deployed in power grid wildly [7]. WSNs are applied through the entire process of the smart grid i.e. from the generation, transmission and distribution, and the consumer side. Some of the applications include load management and control, wireless automatic meter reading (WAMR), equipment fault diagnostic, remote monitoring, fault detection, Advanced Metering Infrastructure (AMI), and residential energy management. In [8] an in depth review on the application characteristics and traffic requirements of several emerging smart grid applications is performed. In order to meet such requirements utilities from all over the word are now facing the key challenge of finding the most appropriate technology. Today, the most widely used protocol for communicating power grid measurement data is IEEE C37.118 that can be transmitted over TCP, UDP or higher level protocols. The C37.118 standard is usually implemented as a client/server communication protocol, where the Phasor Measurement Unit acts as server and the Phasor Data Concentrator as a client. QoS-enabled IP networks are the most suitable communication technology for ensuring a resilient transmission of smart grid monitoring data from the PMUs to the PDC. Fig. 1 shows a view of a Wide Area Monitoring Infrastructure for power grid.

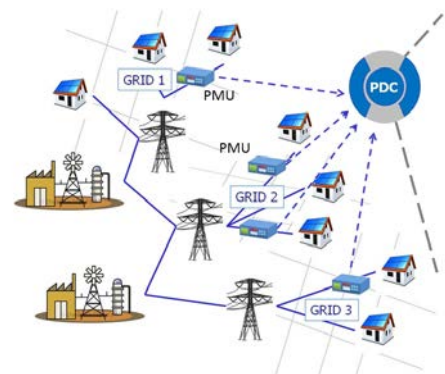

Fig. 1. A schematic view of a power grid monitoring infrastructure

The time synchronization between different PMUs is required to understand the global status of the power grid at the same time. This is because events occurring in one part of the grid affect operations elsewhere, and they also extend to other systems beyond the grid that rely on stable power. Time synchronized measurements produced by PMUs are called synchrophasors. In order to obtain simultaneous measurements of phasors detected from different PMUs installed across a wide area of the power system, it is necessary to synchronize these times, so that all phasor measurements belonging to the same time are truly simultaneous. Each PMU uses a Global Positioning System (GPS) receiver to take a unique timestamp within the global system. One of the main problems affecting smart grid monitoring is the spoofing of the GPS signal provided to the GPS receiver. The GPS signal can be forged in order to mislead the GPS receiver that uses it [9].

In the following we first discuss the security issues introduced by the adoption of the above mentioned networking technologies and then we present the way the MASSIF SIEM system can be used to protect the CI against the discussed cyber-security threats.

*A. Protecting the WSN trunk*

A well-known attack that can be launched by an intruder node against a WSN is commonly referred to as Sleep Deprivation Attack. This attack can be launched in a variety of ways depending on both the particular routing protocol and its specific implementation [10]. As an example it is possible to send many broadcast routing packets. The attack is amplified if the fake routing packets force some nodes to change their parents, since in this case each fooled node notifies the detected change to all its neighbours, thus generating more traffic. Another way to conduct the attack is by sending unnecessary routing requests (RREQ), or by sending forged routing reply (RREP) packets that force the creation of loops in the WSN. In this case, due to the loops, packets are forwarded and stay alive longer, hence resulting in unnecessary retransmissions and additional routing messages. The attack has two negative effects on the WSN: i) the discharge of batteries of all the nodes along the route (the path identified by triple arrows in Fig. 2) from the malicious node to the base station; and, ii) a Denial of Service for those nodes whose path towards the base station (identified with the symbol "x" in Fig. 2) crosses the attacked overloaded path to the base station.

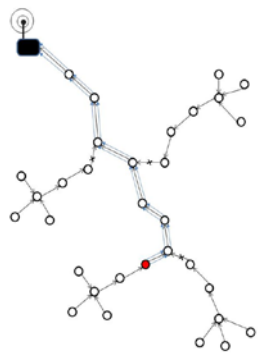

Fig. 2. Sleep deprivation attack and its effects

In a realistic scenario, all packets reaching the base station are typically forwarded to a proxy, which in turn forwards them, on a TCP/IP channel, to the application server for final delivery to the real consumer (Fig. 3). Every data packet generated by a node and reaching the base station is encapsulated into the payload of a TCP packet and sent to the application server via a VPN. In case of an attack, all packets sent by the malicious node will reach the application server which will recognize them as valid packets (as they are duplicates of valid packets), thus resulting in a manipulated view of the field.

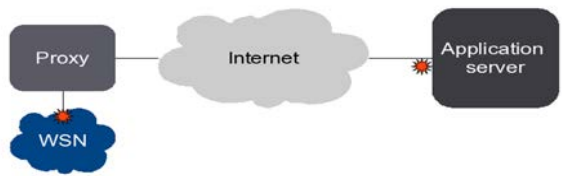

Fig. 3. Deployment scenario for the WS

As an example in our power grid scenario, the attacker may prevent some nodes from sending measures to the related collector, thus hiding changes in the power grid conditions. In order to detect this WSN targeted attacks, we developed a SIEM Correlation Engine capable of correlating alarms given by a WSN security probe and results given by an application level probe used to protect the visualization server. The WSN security probe generates alarms based on the analysis of the network and periodically (e.g. every minute) calculates the packet generation rate at every node. Such data are sent to the correlation module that analyzes them together with alarms triggered by the application level probe for packet arrival rates at the application server higher than a threshold value defined in the training stage. By correlating the data gathered by these two kinds of probes it is possible to detect the sleep deprivation attack and even to identify the ID of the malicious node. The identity of the malicious node can hence be taken into account while applying remediation and recovery actions. The SIEM system can in fact trigger a reaction strategy to stop or at least mitigate the effect of the ongoing attack. Strategies might be selected from a list of possible options, possibly reordered based on their effectiveness with respect to the diagnostic results (e.g., attacker ID, level of damage, class of attack), and sequentially applied until the attack is stopped. Examples of possible reaction strategies for the described attack are (sorted by level of effectiveness): (i) over the air re-programming of the malicious node, (ii) switch-off (i.e., setting to sleep mode) of the malicious node, and (iii) isolation of the malicious node and if necessary manual shutdown and re-programming of the node.

*B. Protecting the wired network trunk*

In this section we illustrate how the MASSIF SIEM system can be used to protect an IP-based network segment composing the power grid monitoring infrastructure against Distributed Denial of Service (DDoS) attacks. Usually, a DDoS attack abuses network protocols in order to saturate the resources of a network server, thus preventing legitimate users from using the provided service. Attackers typically attempt to saturate the limited resources of the victim without directly violating it. A common way to perform a DDoS attack is to first of all compromise other hosts which the attacker will eventually orchestrate during the actual distributed attack. A typical DDoS attack scenario involves several components widely distributed throughout the network: a "master" that initiates and orchestrates the distributed attack, several "agents", or "zombies", which receive commands from the master and launch the real attack, and a target node, which represents the victim of the attack. We herein assume that the attacker perpetrates a DDoS attack against a web based component of a CI (such as the power grid monitoring application server). In this scenario the attacker firstly needs to "recruit" the necessary computing power. Indeed, the greater the available computing power, the stronger the effects of the attack, since the capability of the attacker to saturate the victim's resources increases. For this reason, the attacker needs to opportunely violate several hosts, in order to recruit them as "accomplices". There exist several ways of compromising agent nodes. For example, a Trojan horse might be exploited, allowing the attacker to download an agent on the violated machine. Weaknesses in the software code that accepts remote connections can also be exploited for the recruitment process. Once the victim has been identified, the attacker launches the attack by running the attack code on the master host. The master communicates with agents in order to instruct them about the attack to perform against the target by means of common packet flooding procedures. The distributed flooding is properly orchestrated by the master in order to amplify the effects of the DDoS attack. Let us assume that each agent perpetrates a SYN FLOODING attack to the victim host. SYN FLOODING is a well-known type of attack, wherein the agent sends a succession of TCP SYN requests to the victim without completing the tree-way handshake process with the expected ACK message. Since the victim reserves memory for handling the connection associated with every TCP synchronization request, such attack can rapidly saturate the victim host's resources if the SYN request sending rate exceeds the threshold defined by the timeout mechanism for the synchronization process on the server site. The detection process is performed by monitoring the system under attack at two different levels: a detection module analyzes traffic metrics and compares their values with specific patterns of activities, and alerts the SIEM system if anomalies are detected; another detection module monitors the target machine by controlling operating system parameters and identifies anomalous values of such parameters, which are symptoms of the ongoing attack. The SIEM system correlates these two classes of symptoms and spots a SYN FLOODING attack. Then, it sends an alarm to the reaction component. The level of confidence of the decision-making process depends on the seriousness of the symptoms as well as on the presence of both classes of symptoms. Network monitoring probes are disseminated throughout the network in order to effectively observe the evidence of the distributed attack. The remediation process is initiated with the reception of an alert event. We assume that the PMU and the PDC are connected through a wired network exploited by both legitimate users, and by attackers performing their malicious actions. The attackers target a web server, and attack packets flow through one of the routers shared with the trunk between PMU and PDC as well. Due to the effects of the ongoing attack, the intermediate shared router is not able to satisfy all incoming requests any longer. This affects also the power grid monitoring packets which are no longer forwarded by the compromised router, resulting in loss of monitoring and control data. The reaction strategy might consist in a routing mechanism that allows the communication infrastructure to be resilient to both node/link failures and attacks. The basic assumption is that packets belonging to a single flow can be split at the edge of the network and sent through the core infrastructure along two or more node-disjoint paths. The proposed routing technique, called 'splitting', relies on the possibility to engineer the network traffic by explicitly routing flows (or even parts of them) and can be effectively realized in MPLS (Multi-Protocol Label Switching) networks [11] [12]. Given the above assumptions, standard resilience techniques like, e.g., re-routing of label switched paths, can be effectively applied in case the SIEM system detects an attack: alternative paths can be calculated and the traffic which is suffering from the presence of attack packets can be promptly re-routed along safe links. In the depicted scenario, the presence of splitting helps both preserve the control traffic from being impaired by the attack flows and allow for the prompt application of backup reaction strategies. As an example, in case the smart grid data traffic is routed along link-disjoint paths, the path including the attacked router might be disabled and the portion of smart grid traffic that it was serving might be immediately re-routed along the already available safe path. Thanks to re-routing the attack does not have a strong impact on the performance experienced by the smart grid data traffic.

### C. Protecting the power grid infrastructure from GPS spoofing attack

PMUs are devices that use GPS signals as a common time source and analyze the waveforms of different transmission lines at different locations across a wide-area system at the same moment. In particular they perform a sampling of the waveforms provided by transmission lines and generate the phasors. These phasors are timestamped using the same clock provided by the GPS receiver. These synchronized phasors are called synchrophasors. Such timestamps can be used to compare collected synchrophasors with microsecond precision. In fact, the PDC gathers the data provided by different PMUs and it performs a comparison between the synchrophasors to assess the status of power grid. A PDC can exchange phasors with PDCs at other locations to perform a wide are monitoring. Civilian GPS signals are known to be susceptible to spoofing attacks which make GPS receivers in range believe that they reside at locations different than their real physical locations. In [13] the authors present several GPS spoofing attack detection techniques. Such techniques rely on signal strength and use a threshold-based approach to detect GPS spoofing attacks. A possible technique consists in monitoring the absolute GPS signal strength, i.e. the observed signal strength is compared to the expected one. If the absolute value of the observed signal exceeds some preset threshold, then an alert is raised. Another technique is to monitor the relative GPS signal strength. An extremely large change in relative signal strength could be a symptom of an ongoing attack trying to generate a counterfeit GPS signal to override the true satellite GPS signals. If the signal increases beyond some preset threshold, an alarm would be generated. An extension of the above two techniques implies that the relative and absolute signal strengths are tested individually for each of the incoming satellite signals. The basic assumption is that t if the signal characteristics are too perfect, there is probably something wrong and an alert should be issued. A further technique consists in monitoring satellite identification codes and number of satellite signals received. Many commercial GPS receivers display satellite identification information, but do not record this data or compare to previously recorded data. Keeping track of both the number of satellite signals received and the satellite identification codes over time may prove helpful in determining if foul play is occurring. The proposed SIEM system can be used to detect GPS spoofing attacks since it includes three security probes, each implementing one of the first three above mentioned detection techniques. The alerts raised by the security probes are sent to a correlation engine which processes and correlates the received alarms and then makes the final decision on whether the detected anomalies are symptoms of an ongoing attacks or not.

## V. CONCLUSIONS

This paper addresses a key issue of current and future CIs, namely effective protection of the heterogeneous network infrastructure on which they rely. The paper makes three important contributions. First, it provides a detailed treatment of the security issues resulting from the adoption in CIs of heterogeneous network solutions, and specifically: Wireless Sensor Networks, satellite networks, and QoS-enabled IP connections. Second, it proposes techniques for enhancing current SIEM technology, by improving its capability of detecting and mitigating attacks targeting the heterogeneous network infrastructure of a CI. Third, it shows how these techniques can be implemented with respect to a challenging real world example, specifically the power grid monitoring infrastructure.

## ACKNOWLEDGMENT

The research leading to these results has received funding from the European Commission within the context of the Seventh Framework Programme (FP7/2007-2013) under Grant Agreement No. 313034 (Situation AWare Security Operation Center, SAWSOC Project). It has been also partially supported by the TENACE PRIN Project (n. 20103P34XC) funded by the Italian Ministry of Education, University and Research, and by the "Embedded Systems in critical domains" POR Project (CUP B25B09000100007) funded by the Campania region in the context of the POR Campania FSE 2007-2013, Asse IV and Asse V.